# Effective Charged Exterior Surfaces for Enhanced Ionic Diffusion through Nanopores under Salt Gradients

Long Ma,[1,2,3,#] Xuan An,[1,4,#] Fenhong Song,[4] and Yinghua Qiu [1,2,3,5]*

1. Key Laboratory of High Efficiency and Clean Mechanical Manufacture of Ministry of Education, National Demonstration Center for Experimental Mechanical Engineering Education, School of Mechanical Engineering, Shandong University, Jinan, 250061, China

2. Shenzhen Research Institute of Shandong University, Shenzhen, Guangdong, 518000, China

3. Suzhou Research Institute, Shandong University, Suzhou, Jiangsu, 215123, China

4. School of Energy and Power Engineering, Northeast Electric Power University, Jilin 132012, China

5. Key Laboratory of Ocean Energy Utilization and Energy Conservation of Ministry of Education, Dalian, Liaoning, 116024, China

# These authors contributed equally.

*Corresponding author: yinghua.qiu@sdu.edu.cn

**Abstract**

High-performance osmotic energy conversion requires both large ionic throughput and high ionic selectivity, which can be significantly promoted by exterior surface charges simultaneously, especially for short nanopores. Here, we investigate the enhancement of ionic diffusion by charged exterior surfaces under various conditions and explore corresponding effective charged areas. From simulations, ionic diffusion is promoted more significantly by exterior surface charges through nanopores with a shorter length, wider diameter, and larger surface charge density, or under higher salt gradients. Effective widths of the charged ring regions near nanopores are reversely proportional to the pore length and linearly dependent on the pore diameter, salt gradient, and surface charge density. Due to the important role of effective charged areas in the propagation of ionic diffusion through single nanopores to cases with porous membranes, our results may provide useful guidance to the design and fabrication of porous membranes for practical high-performance osmotic energy harvesting.

**TOC Graphic**

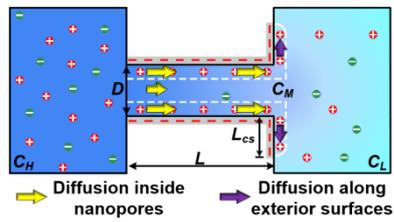

Exploiting sustainable energy is becoming an effective solution to the urgent energy crisis in human society.[1] As a kind of important clean and renewable energy, osmotic energy also known as blue energy, is stored widely between two water bodies with different salt concentrations, such as river water and seawater at the estuaries.[2] With ionic selective porous membranes, osmotic energy can be harvested as electric power through nanofluidic reverse electrodialysis.[3-7] During the energy conversion process, ionic selectivity of nanopores resulted from the electrostatic interaction between surface charges and free ions,[8] enables the selective transport of counterions through nanopores under salt gradients which induces considerable membrane potentials.[2] Osmotic energy conversion (OEC) has deserved much attention due to the gradually increased output power density of porous membranes.[9, 10]

Research on OEC originated from the investigation of ionic diffusion processes through long nanopores and nanochannels.[11, 12] In those cases, the ionic selectivity is determined mainly by the charged inner-pore surfaces.[8] While, charged exterior membrane surfaces have seldom been considered.[13-15] During the OEC process, the output osmotic power depends on both ionic selectivity and permeability,[16] which are mainly controlled by the surface charge property, and membrane thickness, respectively. Because ionic permeability is reversely proportional to the pore length,[17] various thin nanoporous membranes have been fabricated, such as polymer,[18, 19] graphene,[20] MXene,[21] $MoS_2$,[9] and BN membranes,[22] which exhibit ultra-high OEC performance. Feng et al.[9] even obtained the output power density as high as ~$10^6$ W $m^{-2}$ with individual nanopores created on single-layer $MoS_2$ sheets.

Based on theoretical investigations, exterior surface charges can modulate ionic transport through short nanopores, especially for those with sub-200 nm in length.[16, 23, 24] With oppositely charged outer membrane surfaces, significant ionic current rectification was induced through ultra-thin nanopores because of the different ionic selectivity on both membrane sides.[24-26] In conical nanopores with 100 nm in length, exterior surface charges on the tip side can cause obvious ionic current rectification due to the formation of ionic enrichment and depletion zones inside the nanopore under opposite biases.[23] For the osmotic energy conversion, Cao et al.[27] found that the diffusion current and membrane potential were mainly affected by the outer surfaces of short pores. In our previous work,[16, 28] the influences from individual charged surfaces of nanopores on the OEC performance had been systematically investigated under natural salt gradients. We found that electric double layers (EDLs) near charged exterior surfaces on the low-concentration side provide a fast-speed passageway for the counterion diffusion which enhances the ionic selectivity and permeability simultaneously.

The charged exterior surfaces near individual nanopores may be important in the propagation of the OEC performance from single nanopores to porous membranes.[16, 22, 28-30] The total area of charged exterior membranes is determined by porosities. As shown by our earlier work,[16, 28] with the width of charged ring regions around the nanopore ($L_{cs}$) increasing, the diffusion current through nanopores is enhanced which is saturated at a maximum value with fully charged exterior surfaces. The value of $L_{cs}$ at which diffusion current reaches 95% of the maximum current is defined as the effective charged width, which means the smallest required charged area. This idea has been confirmed by a recent experimental study with porous BN membranes.[22] The pore-to-pore distance of 500 nm proposed in nanofluidic experiments agrees well with our simulation results i.e. $L_{cs}$ of ~250 nm for nanopores with 30 nm in length and 10 nm in diameter.[16]

Here, simulations were conducted to investigate the enhanced diffusion current by charged exterior surface and explore the effective charged area under various conditions, such as the pore dimensions, surface charge density, as well as salt gradients and types. From our results, the promotion of ionic diffusion is more significant with a shorter pore length, wider pore diameter, higher salt gradient, or larger surface charge density. The effective charged width $L_{cs}$ is found to be reversely proportional to the pore length, as well as linearly correlated to the pore diameter, salt gradient, and surface charge density, which has a weak dependence on salt types. We think our work can uncover important details of ionic diffusion through nanopores, which enriches the microscale understanding of nanofluidics,[31-33] and provide useful guide to the design of high-performance porous membranes for osmotic energy harvesting.[3, 7]

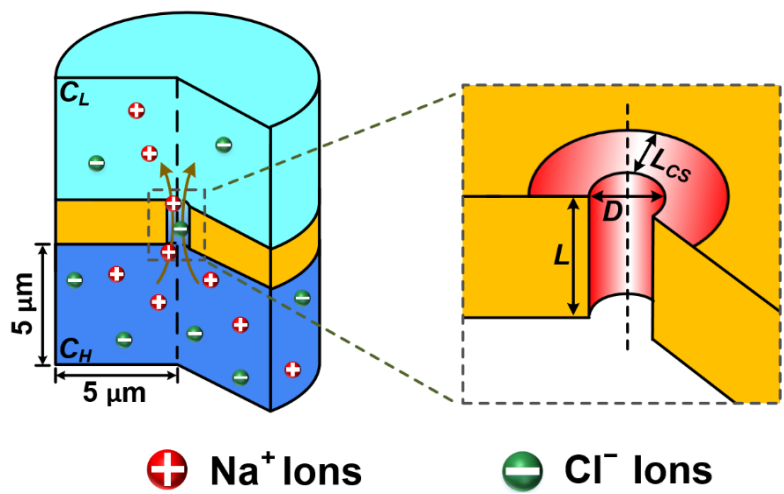

Figure 1 Scheme for nanofluidic simulations under salt gradients. Dark and light blue regions represent two reservoirs with the high- ($C_H$) and low-concentration ($C_L$)

solutions. The yellow part is the membrane with a nanopore of $L$ in length and $D$ in diameter located between both reservoirs. On pore walls, the red zones represent the charged area, and its width is defined as $L_{cs}$.

Ionic diffusion across nanopores under concentration gradients was investigated with three-dimensional simulations through COMSOL Multiphysics. Ion distributions near charged surfaces, ionic transport, and fluid flow through confined spaces were simulated with coupled Poisson-Nernst-Planck (PNP) and Navier-Stokes (NS) equations, as listed below equation 1-4.[34-36]

$$\varepsilon \nabla^2 \varphi = -\sum_{i=1}^{N} z_i F C_i \tag{1}$$

$$\nabla \cdot \mathbf{J}_i = \nabla \cdot \left( C_i \mathbf{u} - D_i \nabla C_i - \frac{F z_i C_i D_i}{RT} \nabla \varphi \right) = 0 \tag{2}$$

$$\mu \nabla^2 \mathbf{u} - \nabla p - \sum_{i=1}^{N} (z_i F C_i) \nabla \varphi = 0 \tag{3}$$

$$\nabla \cdot \mathbf{u} = 0 \tag{4}$$

where $i$ denotes the K$^+$ or Cl$^-$ ions, and $\mathbf{J}_i$, $C_i$, $D_i$ and $z_i$ are their corresponding ionic flux, concentration, diffusion coefficient, and valence, respectively. $\mathbf{u}$ and $\varepsilon$ represent the viscosity and dielectric constant of solutions. $p$, $F$, $R$, $T$, $N$, and $\varphi$ are the hydrostatic pressure, Faraday's constant, gas constant, temperature, number of ion species, and electrical potential, respectively.

From the schematic illustration shown in Figure 1, simulation systems compose two reservoirs and a cylindrical nanopore which forms a sandwich layout. The radius and height of reservoirs were chosen to 5 μm, which is large enough to void the size effect on the ion transport.[37] To figure out the dependence of the effective charged width ($L_{cs}$) of ring regions around nanopores on pore parameters and solution conditions, we have considered different pore lengths, diameters, and surface charge properties, as well as salt types and concentration gradients. Dimensions of the nanopore varied from 2 to 20 nm in diameter, and from 5 to 1000 nm in length. The surface charge density was set from −0.02 to −0.16 C/m$^2$. NaCl, KCl, LiCl, and KF solutions were used to evaluate the influences from diffusion coefficients of cations and anions. Keeping the low concentration at 10 mM, the high concentration varied from 50 to 1000 mM, which provides us with a series of salt gradients from 5 to 100. In our simulation, the default pore dimensions and surface charge density are 10 nm in diameter and 30 nm in length, and −0.08 C/m$^2$,[16, 27, 29] respectively. The natural salt gradient between 500:10 mM NaCl solutions was used for most cases. Detailed boundary conditions[16, 23, 28] and ionic diffusion coefficients[38] were listed in Table S1 and S2. Figure S1 shows the meshing strategy, which can be found in our previous works.[16, 28]

Please note that: based on our previous work,[16, 28] exterior surface charges on the high-concentration side have negligible influences on ionic diffusion due to the strong screening effect of counterions. In order to reduce the computing time, simulation models with only charged exterior surfaces on the low-concentration side were chosen because of the dense meshes on charged surfaces, and such many cases considered in this work.

Diffusion current $I$ was obtained by integration of the total ion flux ($\mathbf{J}_i$) over the reservoir boundary through equation 5. The cation transfer number ($t_+$) was defined from equation 6 with the cation current ($I_+$) and anion current ($I_-$). With $0.5 < t_+ < 1$, the diffusion current of cations is larger than that of anions, and the nanopore is selective to cations.[8]

$$I = \int_S F\left(\sum_i^2 z_i \mathbf{J}_i\right) \cdot \mathbf{n} \, dS \tag{5}$$

where $S$ represents the reservoir boundary, and $\mathbf{n}$ is the unit normal vector.

$$t_+ = \frac{|I_+|}{|I_+| + |I_-|}, 0 \leq t_+ \leq 1 \tag{6}$$

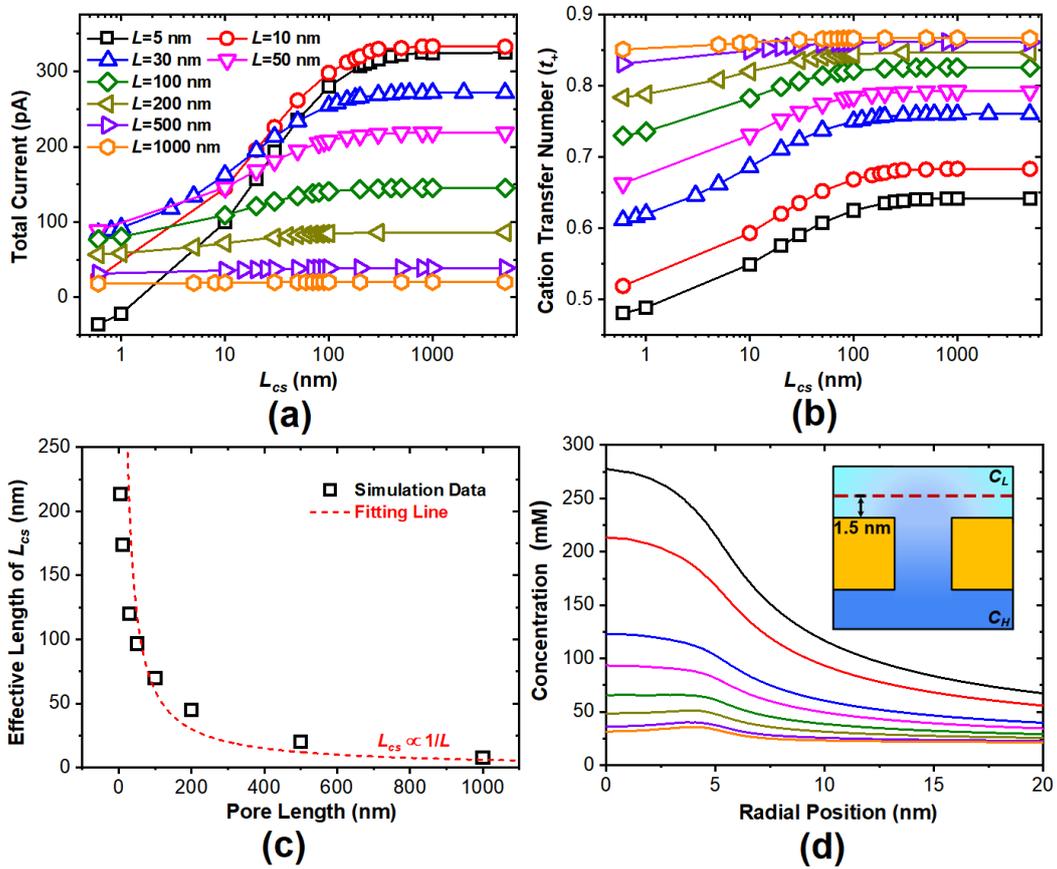

Figure 2 Enhancement of ionic diffusion by the width ($L_{cs}$) of charged exterior surfaces at different pore lengths ($L$). (a) Enhanced diffusion current, and (b) cation transfer number ($t_+$). (c) Effective charged width ($L_{cs}$) for nanopores with different lengths. (d) Concertation distributions of Na$^+$ ions in the radial direction obtained in the plane with 1.5 nm away from the pore exit under $L_{cs}$ = 0. The inset shows the scheme. This plane locates at about the center of EDLs near charged exterior surfaces on the low-concentration side. NaCl solutions on both sides of the nanopore were 0.01 M and 0.5 M, respectively. The pore diameter and surface charge density were set to 10 nm and −0.08 C/m$^2$.

Figure 2 shows the variation of ionic behaviors through 10 nm diameter nanopores with the charged width ($L_{cs}$) on exterior surfaces under different pore lengths. The pore diameter is chosen as 10 nm, which is conveniently obtained in experiments.[39-41] As stated in our previous work,[16, 28] the surface charges carried by the outer wall on the low-concentration side can provide electrostatic interaction to counterions. The formed EDLs regions near the charged exterior membrane surface expand the diffusion area (Figure 3), which induces both promoted ionic selectivity and permeability of counterions.

From Figure 2a, in the nanopore with a length varying from 5 to 1000 nm, diffusion current is enhanced by the increase of $L_{cs}$ until gets saturated. In the cases with shorter nanopores, the promotion to diffusion current from exterior surface charges becomes more significant. For the pore with 5 nm in length, because of the weak electrostatic interaction from surface charges to free ions, Cl$^-$ ions with a higher diffusion coefficient dominate the diffusion process. As the exterior charged area expands, besides the enhanced diffusion current, the nanopore becomes selective to counterions and presents a positive diffusion current. With the pore length increasing, larger inner-pore walls enable the nanopore with a higher selectivity to counterions which causes a stronger diffusion current at $L_{cs}$=0. However, the enhancement by exterior surface charges to ionic diffusion becomes weaker gradually in longer nanopores, exhibiting a gentler slope in the diffusion current increase with $L_{cs}$.

Because of the promotion and suppression from outer surface charges to the diffusion of counterions and coions, ionic selectivity can be strengthened with $L_{cs}$ until reaching stable values. Based on the highly dependent ionic selectivity on the charged inner-pore wall, as shown in Figure 2b, the enhancement by $L_{cs}$ to ionic selectivity is more obvious for thin nanopores and becomes negligible for long nanopores. The maximum increase of $t_+$ is ~33.5% at the pore length of 5 nm. After the pore length approaches the micrometer scale, due to the dominated influences of charged inner surfaces,[16] the modulation from outer charges to ionic diffusion is negligible, resulting in a constant $t_+$ ~0.85.

Energy conversion efficiency is an important parameter to evaluate the performance of osmotic power generation, which depends directly on the ionic selectivity of the nanopore.[11, 12, 16] Due to the reduced charged area near nanopores at a high porosity, porous membranes cannot maintain high selectivity to counterions which results in the deterioration of osmotic energy conversion efficiency.[27, 42] The shared charged area among nanopores induces the competition[43, 44] of ionic diffusion through different nanopores, which may further decrease the energy conversion efficiency.

Based on the above ionic behaviors through nanopores, the effective values of $L_{cs}$ are explored, which may help the quantitative understanding of ionic diffusion through nanopores and provide a useful guide for the design of porous membranes.[16, 22, 28] The effective $L_{cs}$ is determined when its corresponding diffusion current reaches 95% of the current with fully charged exterior surfaces (Figure S2). For nanopores with different lengths, the effective values of $L_{cs}$ is plotted in Figure 2c. We found that the effective $L_{cs}$ is inversely proportional to the pore length under natural salt gradients with 500:10 mM NaCl solutions.[2, 3, 5]

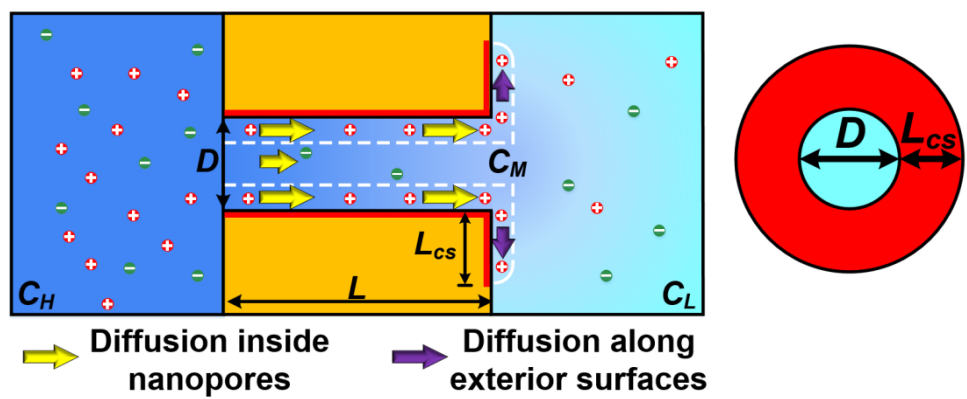

Figure 3 Illustration of ionic diffusion through nanopores. The diffusion process of counterions through nanopores can be divided into two procedures: diffusion inside the nanopore and diffusion along charged exterior surfaces as shown by the yellow and purple arrows, respectively.

From the ionic transport schemed in Figure 3, $Na^+$ ions as the main current carriers diffuse from the high-concentration side to the low-concentration side mainly through the EDLs regions near charged surfaces driven by the concentration gradient. While $Cl^-$ ions as coions pass through the nanopore in the center region. Here, we focus on the diffusion process of counterions which can be divided into two steps, i.e. diffusion inside nanopores and diffusion along exterior surfaces. The effective concentration of counterions at the nanopore exit is defined as $C_M$, which is induced by the ionic diffusion inside the nanopore. The concentration gradient between $C_M$ and $C_L$ provides the driving force for the diffusion along the exterior charged surfaces.

Since the concentration gradient across the membranes is fixed to 500:10 mM, the total motivation for ionic diffusion keeps the same for each case with a different pore length. Due to the inverse dependence of ionic permeability on pore length,[17] diffusion current contributed from individual species exhibits a decreasing trend as the pore length increases. At the exit of a longer nanopore, a lower concentration of counterions appears, i.e. a smaller $C_M$ value (Figure 2d). Please note that the concentration profile is obtained in the plane 1.5 nm away from the pore outer surface, which locates at about the center of the EDLs regions near charged exterior surfaces.[16] Under the weaker driving force between $C_M$ and $C_L$, a smaller effective $L_{cs}$ is obtained, which is inversely proportional to the pore length (Figure 2c).

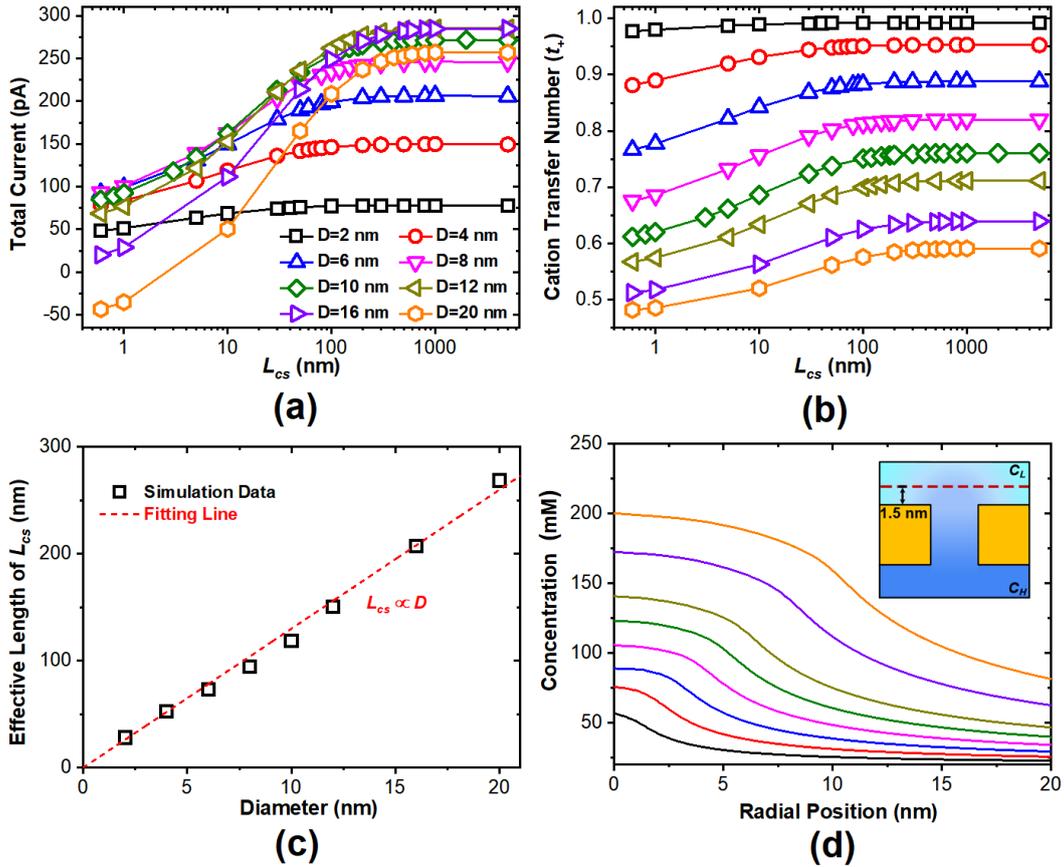

Figure 4 Enhancement of ionic diffusion by the width ($L_{cs}$) of charged exterior surfaces at different pore diameters (*D*). (a) Enhanced diffusion current, and (b) cation transfer number ($t_+$). (c) Effective charged width ($L_{cs}$) for nanopores with different diameters. (d) Concertation distributions of $Na^+$ ions in the radial direction obtained in the plane with 1.5 nm away from the pore exit under $L_{cs}$ = 0. The inset shows the scheme. NaCl solutions on both sides of the nanopore were 0.01 M and 0.5 M, respectively. The pore length and surface charge were set to 30 nm and −0.08 C/m².

During the ionic transport through nanopores, pore diameters determine the cross-section area of the diffusive passageway. As shown in Figure 4a, the enhancement by $L_{cs}$ to the diffusion current is described under different pore diameters varying from 2 to 20 nm. Because of the best performance in osmotic energy conversion for 10 nm diameter nanopores discovered in our previous work,[16] the pore length was set to 30 nm. Following the same trend as that in Figure 2a, the total diffusion current in each case gets promoted with the increase of $L_{cs}$ until archives their maximums. A more significant enhancement in ionic current by $L_{cs}$ happens in the wider nanopores.

In the negatively charged nanopores, ionic selectivity to $Na^+$ ions exhibits a high dependence on the confinement (Figure 4b). In the 2 nm diameter nanopore, the highly confined space induces the overlapping of EDLs near the low-concentration side, which results in an ultra-high ionic selectivity of ~0.98. Although a small diameter can provide higher ion selectivity, it greatly limits the total ionic flux. With the increase of the pore diameter, as illustrated in Figure 3, besides improved diffusion current from $Na^+$ ions, more $Cl^-$ ions can diffuse across the pore in the enlarged center region which results in a decreased ionic selectivity (Figure 4b). When the pore size reaches 20 nm, at a smaller $L_{cs}$ the nanopore is even selective to coions due to its larger diffusion coefficient. This trade-off effect between increased permeability and decreased ionic selectivity leads to an increase-decrease profile of the diffusion current, which reaches a maximum at $D$ ~10 nm (Figure 4a).

The effective $L_{cs}$ on exterior surfaces analyzed under different pore diameters is plotted in Figure 4c, which has a linear relationship to the pore diameter. $Na^+$ and $Cl^-$ ions diffuse across the nanopore through the EDLs regions and pore center, respectively. By defining the local thickness of EDLs in one specified cross-section of the nanopore as $\lambda$ which is determined by the nearby salt concentration,[45] the corresponding cross-section area of EDLs ($S_{EDLs}$) can be roughly evaluated as $S_{EDLs}=\pi\lambda D$. Under the same concentration gradient across nanopores with different sizes, roughly unchanged $\lambda$ locating at the specified cross-section enables the linear correlation of diffusion flux of Na$^+$ ions to the pore diameter. As the increased pore size promotes the transport of counterions, a relatively higher $C_M$ near the pore exit is induced (Figure 4d). Then, the enhanced salt gradient between $C_M$ and $C_L$ results in stronger ionic diffusion parallel to the exterior surfaces which causes a larger effective $L_{cs}$.

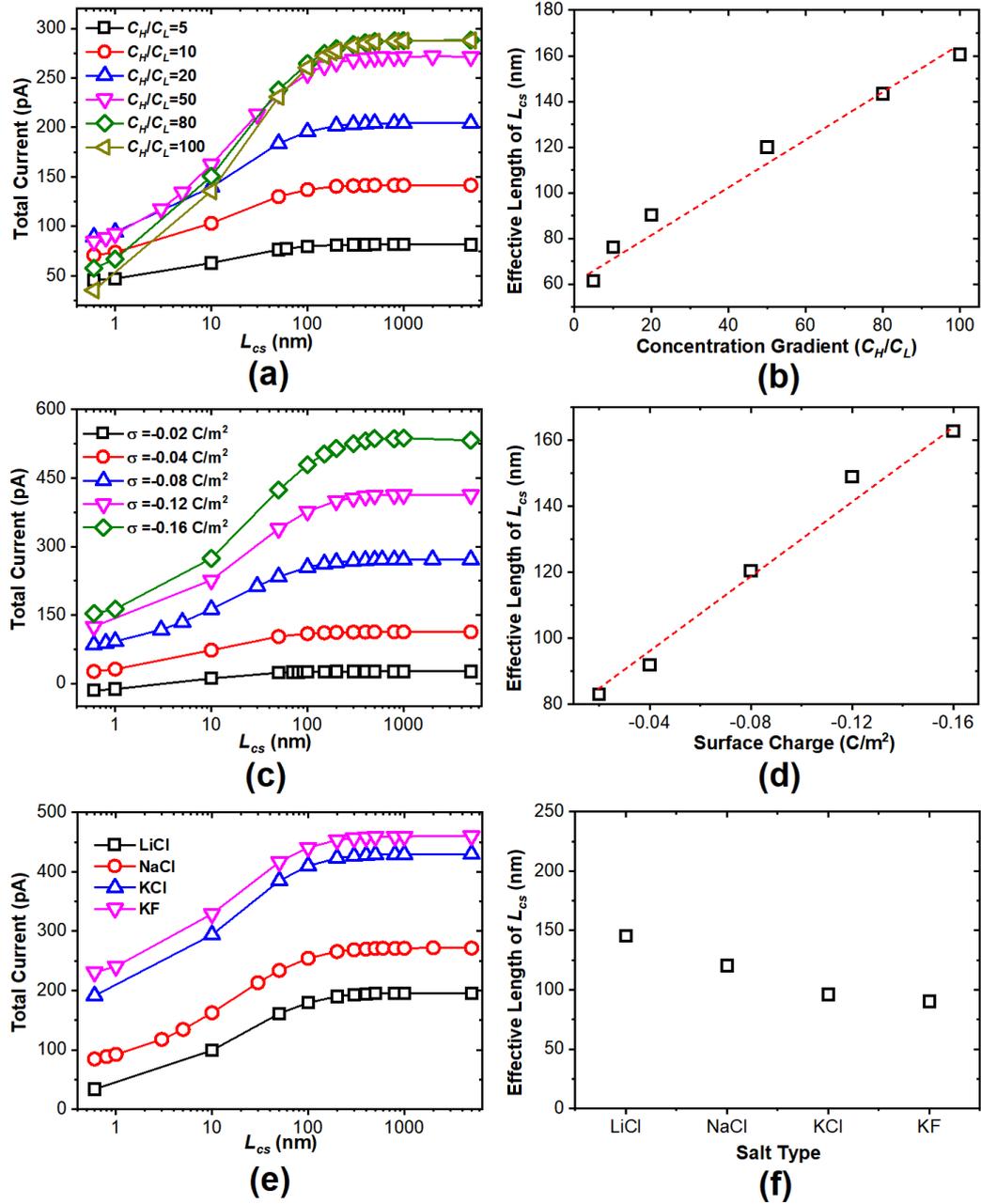

Figure 5 Enhancement of ionic diffusion by the width ($L_{cs}$) of charged exterior surfaces under different conditions, such as concentration gradient, surface charge density, and salt types. Diffusion current (a) and effective $L_{cs}$ (b) under different salt gradients. The low concentration is 0.01 M, and the high concentration is varied from 0.05 to 1 M. Diffusion current (c) and effective $L_{cs}$ (d) under different surface charge densities. Diffusion current (d) and effective $L_{cs}$ (d) obtained with different salt types. In simulations, the pore length and diameter are 30 and 10 nm. Default salt gradient and surface charge density are 500:10 mM NaCl, and −0.08C/m$^2$, respectively.

From the above statement, concentration gradients across nanopores provide the power source for ionic diffusion, and EDLs near charged walls supply a high-speed passageway for counterions transport. Here we explored the enhanced ionic diffusion and effective charged width on the exterior surface under different salt gradients, and surface charge density. Due to the unique diffusion coefficient of different ions, the influence of salt types is also considered. The corresponding ion selectivity of the nanopore in each case is shown in Figure S3.

Following our scheme shown in Figure 3, stronger ionic diffusion appears under a larger concentration gradient because of the larger driving force. While higher concentrations also induce weaker ionic selectivity to counterions (Figure S3a).[46] The trade-off between enhanced ionic flux and weakened ionic selectivity by larger concentration gradients results in an increase-decrease trend in the total diffusion current with smaller values of $L_{cs}$ (Figure 5a). As $L_{cs}$ enlarges, all net diffusion current is enhanced significantly. This improvement is positively correlated to the applied salt gradient. Figure 5b shows the linear dependence of effective $L_{cs}$ on the salt gradient. Because of the unchanged lower salt concentration, increased penetration of counterions enhances the salt concentration $C_M$ at the pore exit (Figure S4a), which induces a larger $L_{cs}$ due to the stronger parallel ionic diffusion to exterior surfaces. As the concentration gradient $C_H/C_L$ increases from 5 to 100, the effective $L_{cs}$ changes from ~60 to 160 nm.

The formation of EDLs originates from surface charges.[46-48] As one important property of nanopores, the surface charge density determines the strength of electrostatic interaction between free ions and charged surfaces, which may inhibit the ionic diffusion process from EDLs regions to locations far away. For more highly charged surfaces, more counterions are accumulated to screen the surface charges, and coions are more depleted due to the larger electrostatic repulsion. Both accumulated counterions in the EDLs regions and depleted coions in the pore center result in higher ionic selectivity (Figure S3b) and net diffusion current, which is also strengthened by the $L_{cs}$. The promotion of diffusion current by charged exterior surfaces is more significant under a higher surface charge density (Figure 5c). As shown in Figure 5d, the effective $L_{cs}$ exhibit a linear relationship to the surface charge density, which is mainly caused by the increased surface diffusion process of counterions (Figure S4b).

The investigation of ionic diffusion through nanopores with different salt types is shown in Figures 5e and 5f. The diffusion coefficient for all considered ions follows the order $Cl^-$ >$K^+$ >$F^-$ >$Na^+$ >$Li^+$ (Table S2).[38] With the increase of $L_{cs}$, diffusion current obtained from all cases is enhanced significantly. The four curves are almost parallel with each other, of which the largest current happens in KF solutions due to the relatively larger diffusion coefficients of counterions. For the cases with the same coions, counterions with a larger diffusion coefficient induce a larger ionic selectivity (Figure S3c) and net diffusion current.[49] Obtained effective $L_{cs}$ is plotted in Figure 5f, which shares similar values varying from ~95 to ~150 nm. We think this phenomenon is mainly resulted from the different diffusion coefficients of counterions. Because of almost the same $C_M$ at the pore exit (Figure S4c), for Li$^+$ ions with the lowest diffusion coefficient, during its transport process along the charged exterior surfaces, the weakest tendency in the diffusion from EDLs regions to locations far away from surfaces brings a relative larger $L_{cs}$. In the cases with KCl and KF, the obtained $L_{cs}$ is almost the same, which is independent of the coions.

In nanofluidic osmotic energy conversion, counterions from the high-concentration side diffuse across the selective porous membrane to the low-concentration side. A considerable membrane potential is induced which can be harvested as electric power. To achieve high-performance energy conversion, nanopores are preferred with high ionic permeability and selectivity, which can be promoted simultaneously by exterior surface charges as discovered in our previous work. Here, simulations are conducted to study the enhancement of diffusion current by charged exterior surfaces with different parameters of nanopores and solution conditions, and explore the effective width of the charged ring regions near the pore. With the appearance of exterior charged surfaces, the diffusion of counterions in EDLs regions can be known as a two-step process: the diffusion inside nanopores and diffusion along exterior surfaces, which are driven by the concentration gradients between the pore entrance and pore exit, as well as between the pore exit and bulk, respectively. Based on our results, exterior surface charges promote the ionic diffusion more significantly through nanopores with a shorter length, a wider size, and stronger surface charge density, or under a higher salt gradient. The corresponding effective charged width exhibits reverse dependence on the pore length, but is linearly correlated to the pore diameter, concentration gradient, and surface charges. Salt types have little influence on this effective value. Explanations of the relationship between effective charged width and various parameters are also provided based on the developed two-step diffusion process. Our finding may not only help us understand the detailed diffusion process across nanopores, but also provide theoretical reference for the production of porous membranes to achieve high-performance osmotic energy conversion.

**Acknowledgment**

The authors thank the support from the National Natural Science Foundation of China (Grant No. 52105579), the Basic and Applied Basic Research Foundation of Guangdong Province (2019A1515110478), the Natural Science Foundation of Shandong Province (ZR2020QE188), the Natural Science Foundation of Jiangsu Province (BK20200234), the Qilu Talented Young Scholar Program of Shandong University, Key Laboratory of High-efficiency and Clean Mechanical Manufacture Ministry of Education, and the Open Foundation of Key Laboratory of Ocean Energy Utilization and Energy Conservation of Ministry of Education (Grant No. LOEC-202109).

**Supplementary material**

See supplementary material for simulation details, and additional simulation results.